\documentclass[prl,twocolumn,showpacs]{revtex4}
\usepackage{ae}
\usepackage{graphicx}

\newcommand{\rb}{$^{87}$Rb}
\newcommand{\rbo}{$^{85}$Rb}
\newcommand{\kq}{$^{41}$K}
\newcommand{\kt}{$^{39}$K}
\newcommand{\kqa}{$^{40}$K}

\newcommand{\ket}[1]{| #1 \rangle}
\renewcommand{\prl}{Phys. Rev. Lett.}
\renewcommand{\pra}{Phys. Rev. A}
\renewcommand{\prb}{Phys. Rev. B}

\begin{document}

\title{Double species condensate with tunable interspecies interactions}

\author{G. Thalhammer, G. Barontini, L. De Sarlo, J. Catani,
  F. Minardi$^{1,2}$ and M. Inguscio$^{1,2}$}

\affiliation{LENS - European Laboratory for Non-Linear
Spectroscopy and Dipartimento di Fisica, Universit\`a di Firenze,
via N. Carrara 1, I-50019 Sesto Fiorentino - Firenze, Italy\\
$^1$CNR-INFM, via G. Sansone 1, I-50019 Sesto Fiorentino -
Firenze, Italy\\
$^2$INFN, via G. Sansone
1, I-50019 Sesto Fiorentino - Firenze, Italy}

\begin{abstract}
  We produce Bose-Einstein condensates of two different species, \rb\
  and \kq, in an optical dipole trap in proximity of interspecies
  Feshbach resonances.  We discover and characterize two Feshbach resonances,
  located around 35 and 79 G, by observing the three-body losses and the
  elastic cross-section. The narrower resonance is exploited to create
  a double species condensate with tunable interactions. Our system
  opens the way to the exploration of double species Mott insulators
  and, more in general, of the quantum phase diagram of the two
  species Bose-Hubbard model.
\end{abstract}



\date{\today}

\maketitle

Ultracold atomic gases seem uniquely suited to experimentally realize
and investigate physics long studied in the domain of condensed matter
and solid state physics, with the distinct advantage that specific
effects are better isolated from unnecessary complications often
present in condensed samples. The paradigmatic superfluid to Mott
insulator transition of a condensate in an optical lattice
\cite{mott-greiner} confirmed the predictions of the Bose-Hubbard
model \cite{bh-fisher,bh-jaksch}, originally introduced to describe
superfluid Helium.  With two species, the zero-temperature diagram of
quantum phases is much richer than the simple duplication of the
single species' \cite{bh2zoo}. Indeed it has been proposed that two
species obeying an extended Bose-Hubbard model can mimick the physics
of lattice spins described by the Heinsenberg model
\cite{spinmap-harvard,spinmap-yale} and give rise to yet unobserved
quantum phases, like the double Mott insulator and the
supercounterflow regime \cite{scf-kuklov}, with peculiar transport
properties. Therefore, a double species condensate in an optical
lattice stands as a promising candidate system for quantum simulations.
Recently, the investigation of the two-species BH was
started from the regime where one species exhibits the loss of phase
coherence usually associated with the Mott insulator transition, while
the other is completely delocalized \cite{bbmix-lens}. Already at this
stage, the presence of two species leads to a surprising shift of the
critical point, which is now object of intense theoretical work
\cite{bbmix-torino}.

In addition, a double Mott insulator is expectedly extremely useful to
produce heteronuclear polar molecules \cite{dipolar-damski}, since the
association efficiency strongly depends on the phase space overlap of
the two species \cite{fesh-efficiency-jila}. The rapid losses of
associated molecules observed for bosonic systems could be largely
suppressed by the presence of the three-dimensional optical lattice
\cite{lattice-molecules-gregor}, if most of the sites are occupied
with only one atom per species.  Both these research avenues require
the dynamic control of {\it interspecies} interactions, along with a
well established collisional model.

In this Letter, we report the production of the first double species
condensate with tunable interspecies interactions. A mixture of two
condensates in different internal states of the same isotope was
realized long ago \cite{doppiabec-cornell} and more recently
Bose-Einstein condensation (BEC) of two different atomic species has
been demonstrated in a harmonic potential \cite{doppiabec-lens,
  doppiabec-jila} and in a three-dimensional optical lattice
\cite{bbmix-lens}. Providing Bose-Bose mixtures with the
additional tool of tunable interspecies interactions available, 
this work meets a long sought goal.

To control the interactions, we investigate two heteronuclear Feshbach
resonances predicted \cite{tieman, simoni}, but yet unobserved, for
both \rb\ and \kq\ in the $\ket{F=1,m_f=1}$ state below 100~G. These
Feshbach resonances are interesting in themselves, since the accurate
determination of the K-Rb potential curves for the singlet
X$^1\Sigma^+$ and triplet a$^3\Sigma^+$ states has been recently
debated \cite{tieman, simoni}. Indeed, extensive Feshbach spectroscopy
on the mixture \rb-\kqa\ was performed by two different groups
\cite{ffr8740-lens,ffr8740-hannover}. Until recently, however, data on
other isotopomers, crucial to pinpoint the controversial number of
bound states of the singlet potential, were not available. Together
with the observation of several Feshbach resonances for the \rb-\kt\
mixture \cite{simoni}, our measurements are important to settle the
question of the number of singlet bound states and to assess the
validity of the mass scaling techniques, based on the Born-Oppenheimer
approximation.

Since our setup has been described earlier~\cite{catani-pra,
  luis-pra}, here we briefly illustrate the experimental procedure. In
separate vacuum chambers, each atomic species, \rb\ and \kq, is coaxed
into a cold atomic beam by means of two-dimensional magneto-optical
traps (2D-MOT's). The two atomic beams merge with an angle of
160$^\circ$ at the center of a third vacuum chamber and load a double
species 3D-MOT. After 50~ms of compressed MOT and 5~ms of optical
molasses, both species are optically pumped to the $|F=2,m_F=2\rangle$
state and loaded into a quadrupole trap with axial gradient equal to
260~G/cm. A motorized translation stage moves the quadrupole coils by
26~mm and atoms are transferred to our Ioffe millimetric trap
\cite{ruquan-pra}. 
For 15.5~s, a microwave field evaporates only \rb, while thermal
equilibrium with \kq\ is enforced by efficient interspecies
collisions.

To access Feshbach resonances we transfer the atoms in a crossed
dipole trap.  At the end of microwave evaporation, we ramp up two
horizontal orthogonal beams of waists $\sim 100\,\mu$m, delivered by a
single-frequency laser at 1064~nm, in 250 ms, and then slowly
estinguish the magnetic trap current.  At this stage we have $3\cdot
10^5$ atoms of \rb\ and $2\cdot 10^4$ atoms of \kq. For \rb\ atoms,
the optical trap has a depth of $7\, \mu$K and harmonic frequencies
equal to $\vec{\omega}\simeq 2\pi\times(100,140,100)$~Hz. Starting
from a thermal single-species \rb\ sample at 350 nK, we measured the
lifetime and the heating rate of our crossed dipole trap to be 20~s
and 20~nK/s.

In order to transfer the atoms to $|1,1\rangle$ state, we apply a
6.8~GHz microwave and a 269~MHz radiofrequency sweep (adiabatic
passage), in presence of a polarizing magnetic field of 7~G: the
transfer efficiency is 90(80)\% for \rb\ (\kq). In the
$|1,1\rangle+|1,1\rangle$ lowest Zeeman state the mixture is stable
against two-body collisions.  Once completed the hyperfine transfer,
\rb\ atoms remaining in $|2,2\rangle$ are expelled from the optical
trap by pulsing a beam resonant with the closed $|2,2\rangle
\rightarrow |3,3\rangle$ transition: their presence would lead to spin
changing collisions with \kq\ releasing the \rb\ hyperfine energy
(0.33 K). For \kq, instead, we omit to expel the remaining
$|2,2\rangle$ atoms, because their absolute number is low and they are
immune from spin changing collisions with \rb\ due to energy
conservation.

\begin{figure}[b]
\centering
\includegraphics[width=\columnwidth]{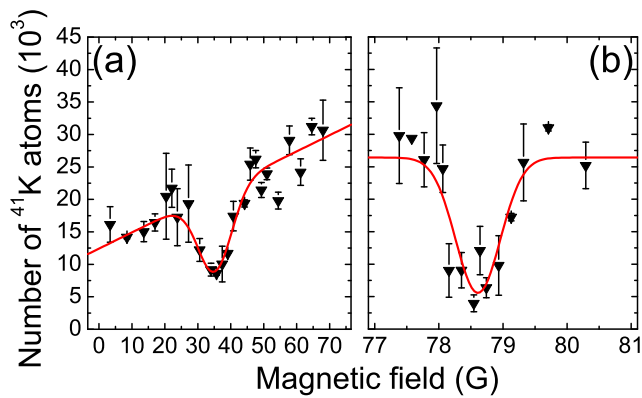}
\caption{(Color online): Losses of \kq\ atoms associated with
  interspecies \rb-\kq\ Feshbach resonances. The solid (red) line
  shows the results of Gaussian fits, yielding the peak values: (a)
  $B_0=(35.2\pm 1.6)$~G, (b) $B_0=(78.61\pm 0.12)$~G. }
\label{fig:3body_losses}
\end{figure}

At this stage, to observe the interspecies Feshbach resonances, we
ramp up the applied magnetic field in 10~ms, hold the atoms for
500~ms, abruptly switch off the optical dipole trap, and image
separately the clouds after expansion times of 5 to 10~ms. We observe
the enhanced three-body losses associated with the Feshbach
resonances, which are best detected as a drop of the number of the
minority fraction, i.e. \kq. To avoid the complications arising from
the dynamics driven by mean-field interactions coupled to the
differential sag between the two species, we detect the losses with
thermal clouds at $\sim 1\,\mu$K.

We scan the magnetic field in the range 0-90~G and we detect two
Feshbach resonances, around 35 and 79~G, see
Fig.~\ref{fig:3body_losses}. We fit the loss features with Gaussian
functions of the magnetic field. For the broader feature around 35~G,
we add a linear pedestal, that takes into account the depolarization
of the \kq\ sample, caused by incomplete suppression of laser light
resonant at zero magnetic field. The loss peaks occur at $(35.2\pm
1.6)$ and $(78.61\pm 0.12)$~G \cite{errorbars} and the widths are
$(5.1\pm 1.8)$ and $(0.35\pm 0.14)$~G, respectively. We calibrate the
magnetic field by measuring the frequency of \rb\ hyperfine
transitions; the associated systematic uncertainty is 0.4~G.

\begin{figure}[b]
\centering
\includegraphics[width=.9\columnwidth]{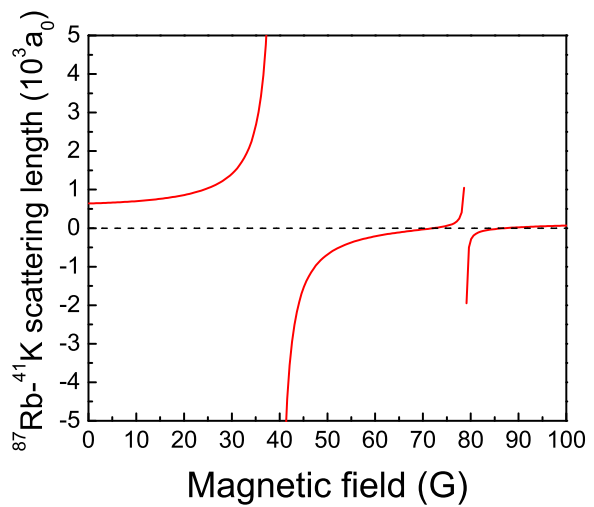}
\caption{(Color online): Theoretical predictions of the \rb-\kq\
  interspecies scattering length $a_{12}$ \cite{simoni}, with both
  species in the $|1,1\rangle$ lowest Zeeman state.}
\label{fig:theo}
\end{figure}

We compare the observed Feshbach resonances positions with the
theoretical predictions, shown in Fig.~\ref{fig:theo}, of the
collisional model of Ref.~\cite{simoni}, yielding resonance positions
equal to $(39.4\pm 0.2)$ and $(78.92\pm 0.09)$~G. In particular, the
position of the narrower loss feature around 79~G agrees very well
with the theoretical value of the Feshbach resonance. For the broader
feature, the measured three-body losses are maximum at a field
slightly below the theoretical prediction. We lack a conclusive
explanation for this difference, we merely remark that the modeling of
the three-body losses is complicated and, as a general rule, for broad
resonances the loss peak can be offset from the resonance position.

In view of the application of these Feshbach resonances to tune the
mixture around the non-interacting regime, e.g. for producing a double
species Mott insulator phase, it is of special interest to locate the
{\it zero-crossings} of the interspecies scattering length $a_{12}$,
i.e. the magnetic field values where $a_{12}$ vanishes. To this end,
we studied the efficiency of sympathetic cooling of \kq\ by \rb\
\cite{control4078-bec2}. For this, we applied an additional
evaporation stage by lowering the optical trap power to 50\% in
1.5~s. Even though the optical potential is 10\% larger for \rb, the
combined optical and gravitational potential makes the optical trap
deeper for \kq. Therefore by lowering the trap depth, we evaporate
mostly \rb\ and rely on sympathetic cooling for \kq.

\begin{figure}[t]
\centering
\includegraphics[width=\columnwidth]{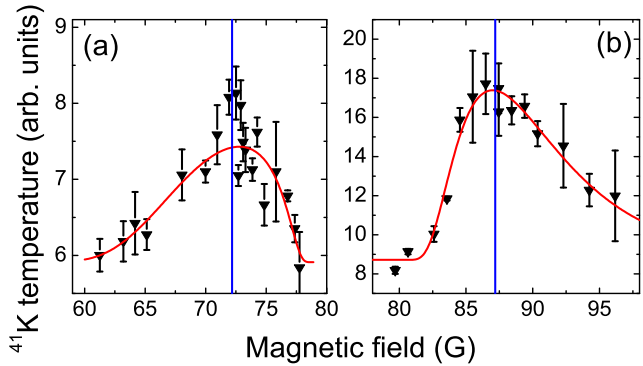}
\caption{(Color online): Measurement of the magnetic field of
  vanishing $a_{12}$ (zero-crossing). The \kq\ temperature after
  sympathetic cooling with \rb\ is fit (red line) with the function
  $T_{\rm eq} + \Delta T \exp(-\eta (a_{12}(B-\delta B))^2)$. $a_{12}$
  as a function of $B$ is derived by the theory, $T_{\rm eq},\Delta T,
  \eta, \delta B$ are fitting parameters. The vertical (blue) lines
  indicate the zero-crossing theoretical predictions: 72.18~G (a) and
  87.19~G (b).}
\label{fig:zero_crossing}
\end{figure}

During the thermalization the temperature of \kq\ decreases
exponentially with a rate proportional to the interspecies elastic
cross section \cite{thermalization}, proportional to $a_{12}^2$, hence
the final temperature of the \kq\ cloud $T_{41}^{\rm fin}$ is a
decreasing function of $a_{12}^2$: $T_{41}^{\rm fin}=T_{\rm eq}
+ \Delta T \exp(-\eta a_{12}^2)$, where $\eta$ is a parameter
depending on the overlap density and the thermalization time. We fit
our data with the above simple model where the dependence of $a_{12}$
on the magnetic field is taken from the theory.  In addition to
$T_{\rm eq},\Delta T$ and $\eta$, we introduce another fit parameter
$\delta B$ allowing for a global offset of the magnetic field values,
to quantitatively verify the agreement of our data with the
theoretical predictions. In Fig.~\ref{fig:zero_crossing}, we show the
data plot along with the fit. Since the $\delta B$ fit values equal
$(0.59\pm 0.64)$ for Fig.~\ref{fig:zero_crossing}(a) and $(-0.21\pm
0.32)$~G for Fig.~\ref{fig:zero_crossing}(b), we conclude that the
zero-crossing measurements are in good agreement with the theoretical
predictions.

With the benefit of the controlled interspecies interactions arising
from such favorable Feshbach resonances below 100~G, we have been able
to produce the double species condensate on both sides of the narrow
Feshbach resonance. For this, we slightly modify the final stages of
evaporation and sympathetic cooling, that must be adjusted for the
targeted value of interspecies scattering length. For example, to
achieve BEC on the repulsive side, we first ramp the magnetic field to
77.7~G in 40~ms, where expectedly $a_{12} \simeq 250a_0$. We then
evaporate the mixture by lowering the optical trap power in 3~s.
During the last 500~ms of evaporation, the magnetic field is brought
to 76.8~G ($a_{12} \simeq 150a_0$), to reduce the rate of three-body
losses.  With this scheme, we produced pure double species condensates
with typically $9(5)\cdot 10^3$ \rb\ (\kq) atoms.
In alternative, we achieve a double condensate, with the same typical
number of atoms, by performing the optical evaporation at 80.7~G,
where $a_{12}\simeq-185a_0$. 

\begin{figure}[b]
\centering
\includegraphics[width=.85\columnwidth]{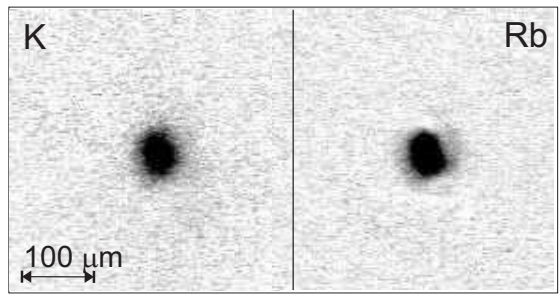}
\caption{Double species condensate: both species are in the $|1,1\rangle$
  state. The image is taken after 9 ms of expansion.}
\label{fig:image_2bec}
\end{figure}

This fact highlights an important feature of our setup. At the end of
evaporation, the measured harmonic frequency of the dipole trap along
the vertical direction equals 84~Hz (for \rb) and the trap centers for
the two clouds are $13\,\mu$m apart, due to the differential gravity
sag. By suitably balancing the number of atoms, we arrange BEC to
occur first for \kq. Shortly afterwards, the two clouds separate and
\rb\ reaches BEC. This separation, that can be reversed by
recompressing the trap at the end of the evaporation, allows us to
avoid all mean-field dynamics induced by interactions between the two
species, like the phase separation(collapse) expected for large
positive(negative) $a_{12}$ \cite{topology-riboli}. At variance,
striking manifestations of mean-field effects were observed in a
recent experiment: in a \rb-\rbo\ double condensate, tuning the \rbo\
scattering length leads to phase-separation and formation of
long-lived droplets \cite{phase-sep-jila}.  Indeed, the solution of
coupled Gross-Pitaevskii equations shows that the mean-field
interactions dictate the equilibrium configuration through the
parameter $\Delta=(a_1 a_2/a_{12}^2)-1$. This parameter can be tuned
by varying either $a_1$ or $a_{12}$. However, for the purpose of
exploration of the phase diagram and molecular association, tuning the
single-species scattering length is not sufficient and the ability of
varying $a_{12}$ is required. 

For this goal our \rb-\kq\ mixture is especially valuable, because to
date it represents the only heteronuclear Bose-Bose mixture where an
interspecies Feshbach resonance is accessible with stable condensates
of both species.  Particularly important is the possibility,
here demonstrated, to precisely control $a_{12}$ around zero. The
observed zero-crossings appear relatively comfortable, as the slope is
respectively of 16.8 and 8.9 $a_0/$G, and we foresee a control on
$a_{12}$ within a precision better than $1a_0$.

In conclusion, we have experimentally observed two Feshbach resonances
at low (<100~G) magnetic fields, for the mixture \rb-\kq\ in the lowest
Zeeman state. These Feshbach resonances allowed us to create the first
double species Bose condensate with tunable interspecies
interactions. Combined with a three-dimensional optical lattice, which
we have already implemented and used \cite{bbmix-lens}, this tool will
enable the exploration of the two-species BH phase diagram. For this
purpose, we measured the position of two zero-crossings of the
interspecies scattering length.  

The Feshbach resonance around 79~G appears well suited to create
molecules by sweeping the magnetic field, due to its narrow width of
1.2~G. On the other hand, the broad resonance at lower magnetic field
might be more favorable for radiofrequency molecular association,
since the energy separation $\Delta E$ of the bound state from the
free atoms threshold increases very slowly with the magnetic field
detuning from resonance $B-B_o$, i.e. $\Delta E \simeq
-0.005\mu_B(B-B_0)$.

Finally, our heteronuclear mixture with tunable interactions will
allow the investigation of the intriguing physics of a superfluid
interacting in a controlled way with a material grating, constituted
by localized bosonic atoms in a deep periodic optical potential. For
such systems, polarons growth and dynamics are currently under
theoretical analysis \cite{polaron-jaksch}.

This work was supported by Ente CdR in Firenze, INFN through the
project SQUAT-Super, and EU under Integrated Project SCALA, Contract
No. HPRICT1999-00111 and Contract No. MEIF-CT-2004-009939. We are
grateful to A. Simoni for sharing his theoretical results prior to
publication. We also acknowledge useful discussions with all members
of the Quantum Degenerate Gas group at LENS.

\end{document}